# OpenDSU: Digital Sovereignty in PharmaLedger

Cosmin Ursache[a], Michael Sammeth[b,c], Sînică Alboaie[a,d]

a) RomSoft SRL, Bulevardul Chimiei 2bis, Iași 700291, Romania {urc,als}@rms.ro
b) University Hospital Würzburg, Oberdürrbacher Str. 6, 97080 Würzburg, Germany Sammeth_M@ukw.de
c) Coburg University, Friedrich-Streib-Str. 2, 96450 Coburg, Germany Michael.Sammeth@hs-coburg.de
d) Axiologic Research SRL, Str Armeana nr 10, Iași 700039, Romania sinica@axiologic.net

**Abstract**

Distributed ledger networks, chiefly those based on blockchain technologies, currently are heralding a next generation of computer systems that aims to suit modern users' demands. Over the recent years, several technologies for blockchains, off-chaining strategies, as well as decentralised and respectively self-sovereign identity systems have shot up so fast that standardisation of the protocols is lagging behind, severely hampering the interoperability of different approaches. Moreover, most of the currently available solutions for distributed ledgers focus on either home users or enterprise use case scenarios, failing to provide integrative solutions addressing the needs of both.

Herein we introduce the OpenDSU platform that allows to interoperate generic blockchain technologies, organised – and possibly cascaded in a hierarchical fashion – in domains. To achieve this flexibility, we seamlessly integrated a set of well conceived OpenDSU components to orchestrate off-chain data with granularly resolved and cryptographically secure access levels that are nested with sovereign identities across the different domains.

Employing our platform to PharmaLedger, an inter-European network for the standardisation of data handling in the pharmaceutical industry and in healthcare, we demonstrate that OpenDSU can cope with generic demands of heterogeneous use cases in both, performance and handling substantially different business policies. Importantly, whereas available solutions commonly require a pre-defined and fixed set of components, no such vendor lock-in restrictions on the blockchain technology or identity system exist in OpenDSU, making systems built on it flexibly adaptable to new standards evolving in the future.

# 1. Introduction: Modern IT Systems

In a world that is becoming more digital than ever before, new technologies are emerging to satisfy the needs of modern users to connect with each other. Recently, the development of *distributed ledger technologies* (DLTs) has fueled a plethora of novel concepts regarding the exchange of information, data and other digital assets (Sec. 2). Technically, most of the successful DLT approaches rely on *blockchains*, which redundantly store and process user transactions in a distributed network. In the remainder of this section, we briefly sketch challenges and chances of these decentralised DLT systems.

*1.1 Decentralised Internet Services*

Especially over the last decade, services for every aspect of life have increasingly become accessible through internet platforms. Already during the commercialisation of the world wide web, infrastructure gradually shifted from "web1", where each user as an equivalent node could produce and/or consume content, to today's "web2" structure [1], where centralised platforms operated by a small group of "Big Tech" companies administrate services required by home users and/or smaller companies [2]. Recently, the concept of a next-generation "web3" architecture promoting the return to decentralisation [3] is increasingly drawing attention.

Although concrete visions of web3 are still varying, a common key concept is that both the ownership and the control of internet services should both be decentralised again. *Decentralised ownership* of exchangeable assets is usually broken down by tokens [4], and non-fungible tokens (NFTs) have been conceived to represent digital counterparts of unique assets such as artefacts, credentials, governance rights, access passes, etc [5]. The *decentralised control* of services is achieved through internet platforms that democratise and streamline user interactions based on DLT networks instead of relying on a centralised database.

Later web3 platforms are commonly known as *decentralised applications* (dApps), e.g. dApps for Decentralised Finance (DeFi) allowing users to exchange currency without any involvement by a bank or government. On the top of such decentralised architecture, web3 supporters further see a novel form of organisations substituting the traditional coordination patterns of corporate business in the long run: these so-called *Decentralised Autonomous Organisations* (DAOs) are internet-based companies that are collectively owned and controlled by their respective members. However, today's platforms are still falling for the web2 centralisation traps, as most of the web3-based concepts currently remain visionary, lacking concrete implementations of the necessary tools.

*1.2 Sovereign Identity Systems*

To access services on internet platforms, users need to transfer their own identity to the digital world. To enable trustful interactions between digital identities, the organisation-centric ID management authenticates users based on passwords that are controlled separately by each data provider in the form of a traditional database, resulting in obvious privacy concerns. Later-on, in the progress of web2, the *federated identity systems* have emerged, allowing users to verify their digital IDs by password-based credentials also with third parties, without exposing personal information to each of them. However, whilst reducing the number of data providers with access to private user data, concerns about potential data leakage and abuse by malicious identity providers can obviously also not be refused in federated systems.

Therefore modern DLT systems are moving towards more user-centric approaches to identity control, like *Decentralised Identifiers* (DIDs) in general allowing for *Self-Sovereign Identities* (SSIs) in particular [4]. Operating without centralised registries, identity providers, and certificate authorities by design, DIDs can be used to represent people but also objects [5], datasets [6], and many more. The World Wide Web Consortium (W3C) recently released a DID recommendation based on ten principles [7] as well as a recommendation on *Verifiable Credentials* (VCs) to promote verification processes and by this to "bind" a certain user to some data or transaction [8].

In this light, VCs are associated with digital identities in the same way physical credentials are tied to traditional identities. However, VCs are designed for web environments, where it is more difficult to verify or validate the information because plain digital patterns are more easily tampered with than their physical-biological counterparts. VCs therefore require to be cryptographically secure, privacy-respecting, and machine-verifiable. Driven by motivations to preserve a maximum of privacy, VCs are postulated to reveal a minimal amount of personal information exposed by the users – a maxim for which the term *Zero-Knowledge Proofs* (ZKPs) has been coined [9]. Employing ZKPs, the issues in an identity system are shifting from the former concerns about potentially malicious data controllers to new concerns about potentially malicious users, bringing up the question of how to create trust in a trustless digital environment. Therefore, a major challenge of introducing SSIs in DLT systems remains the fine tuning of the balance between the opposingly seeming requirements of privacy and trust.

*1.3 Blockchains and Scalability*

A downside of decentralised blockchains is that data on the blockchain (i.e., "*on-chain*" data) and corresponding computations are replicated redundantly by all the nodes in the network. By this, computationally intensive transactions in combination with the performance of the least performant node can slow down the blockchain network, hampering the throughput and also the scalability of the entire system. According to the classification of [10], such "computationally intensive" transactions consist of transactions that include costly computations of deferred results, transactions with accumulating data, or transactions with heavy data files.

Naive attempts to circumvent such performance issues by keeping large files separately on premises beyond the blockchain ("*no-chain*" data) result in a loss of data integrity and auditability, as well as in issues with the protection particularly of sensitive data. Therefore an increasingly explored approach consists in computing hash codes that are kept on-chain before relocating heavy or sensitive files to systems outside the blockchain (*off-chain*" data).

However, importing off-chain data to/from external premises also raises novel challenges: first, the volume of off-chain data often grows manifold faster than on-chain data, e.g. >40-fold in an empirical study on a 5-node Hyperledger system [11]; second, a *centralised* off-chain storage re-introduces previously discussed problems of data control, whereas a *decentralised* off-chain storages questions about security, availability and delivery of the off-chain data; third, off-chain data needs to be validated upon re-import, e.g. by leveraging hashing techniques, imposing additional overhead to the system. In a nutshell, current key challenges for off-chain storage strategies comprise data access regulation, storage security, cryptography, overall performance, and system scalability.

| *Use Case Pattern* | #1 Single Data Provider | #2 Decentralised Choreographies | #3 Trustless Choreographies | #4 Global / Public Networks |
|---|---|---|---|---|
| **Example Use Cases** | NFTs & Data Anchoring | Business Processes | DAOs | Crypto currencies, DeFi systems |
| **Stakeholders** | Single entity sharing data | Business partners updating data | A large group of independent peers | A network of independent parties |
| **Privacy** | Encryption, P2P messaging | Encryption, P2P messaging | ZKP, Encryption, Layer 2 solutions | ZKP |
| **DoS/Spam Prevention** | - | Legal contracts | ZKP, Transaction fees | ZKP, Transaction fees |

*Table 1 - Use Case Patterns*: Use cases from permissioned networks are grouped into four distinct collaboration categories ("use case patterns"), according to characteristics of the stakeholder(s) involved, their privacy requirements and possibilities to prevent spamming attacks.

*1.4 Use Case Patterns*

Ledgers are employed to model a variety of use cases from different business sectors, such as financial accounting, logistics, healthcare, etc. These use cases inherently differ in the number and degree of mutual trust of the users interacting with each other, and consequently also in the challenges for implementing them. Based on our research on best practices for programming ledger solutions, we classify use cases in one of four abstract interaction scenarios we call *use case patterns* (Table 1):

*Pattern #1 (Single Data Provider)* describes use cases, where a single entity is sharing its data with partners. In a typical use case of this category, a – potentially big – company ("producer") is providing controlled read-only access to its NFTs for their partners ("consumers"). Although scenarios along these lines could be solved already by employing traditional databases, blockchain solutions are today often preferred due to the inherent advantages of decentralisation in offering data integrity and audibility. In fact, most enterprise blockchain use cases classify either directly as pattern #1 or can be redesigned in a way that every data item has a single controller. Privacy is usually assured by direct peer-to-peer (P2P) messaging and encryption. Denial-of-Service (DoS) attacks, such as adversary spamming, and Replay Attacks are not relevant to pattern #1 use cases, since the only meaningful attacks could originate from the data controller itself.

*Pattern #2 (Decentralised Choreographies)* comprises use cases within a – usually small – group of business partners who are interchanging their respective data, such as for instance in the banking sector or in multi-centric clinical studies. Since the partners involved in the business processes mutually trust each other, privacy can still efficiently be preserved by encryption and P2P messaging. As in Category #1 use cases, DoS attacks are of subordinate relevance because all transaction partners are bound with each other by legal contracts. This pattern comes as a good solution for the corner cases where pattern #1 does not suffice. For instance, modelling the data stats of a business process between multiple parties represent a very general use case for pattern #2.

*Pattern # 3 (Trustless Choreographies)* describes use cases amongst parties in a – usually larger – group, usually spawning (sub-)clusters of cooperating partners or otherwise not one-to-one connected peers. The fact that the pattern includes several stakeholders hampers straightforward extensions of solutions for use cases from patterns #1 and #2 in their confidentiality protection, and makes them vulnerable to spam attacks. Therefore, technologies like ZKPs should be deployed (Sec. 1.1). A typical use case of pattern #3 could be constituted by a network of public DAOs that have smart contracts but are independent of other DAOs, respectively, they do not need to sync in real-time with the status of the whole network. The natural segregation of the users in the entire network into largely independently operating clusters enables various performance optimisations, that are explored e.g. by the side-, child- and off-chaining approaches in Layer 2 solutions (Sec. 2.3).

*Pattern #4 (Global/Public Networks)* subsumes all use cases in which anonymous parties interact with each other through transactions in a decentralised ledger network. Classical examples are the popular cryptocurrency blockchains (e.g. Bitcoin), DeFis, or other networks with transactions of valuable things. In such scenarios, the whole network has the interest to ensure correctness. As for use cases of pattern #3, ZKPs are required to verify the identity of anonymous entities in such public networks. The subtle difference between the patterns #3 and #4 is that in pattern #3 some off-chain data integrity can be tolerated without serious risks, and on-chain data integrity can be minimised but not entirely removed.

## 2. Related work

A vast variety of DLTs has been developed over the recent years, however, comparatively little effort has been spent on the systematic classification, evaluation and standardisation of different approaches. Consequently, today's landscape of ledger solutions is marked by a high degree of diversification in their concept of smart contracts and communication patterns, originating also from the heterogenous challenges of the use cases for which they were designed. Focusing on permissioned ledger technologies, we subsume in the remainder of this section some popular DLTs that are related to our work [12].

*2.1 UTXO models in public ledgers*

Going back to the famous Bitcoin system [13], global ledgers have been conceived principally to address the use cases described as pattern #3 (permissioned / private) and #4 (unpermissioned / public) in Tab. 1. A common attribute of global ledgers is their use of a global transaction model called UTXO (unspent transaction output). Under this model, UTXO represents the current state of a global database, from which transactions can "spend" or "consume" assets as inputs, and corresponding outputs then are incorporated in the updated UTXO database. In contrast to ledgers employing account models (Sec. 2.2), smart contracts in UTXO systems are essential for guaranteeing global data integrity, checking the validity of every submitted transaction with respect to its correctness, authorisation and uniqueness.

**Corda: Smart Contracts and P2P messaging**

Developed mainly by the company R3 with other partners involved, Corda[1] is an open-source project that (inter-)operates permissioned networks with Smart Contracts executed in a JVM-compatible language (i.e., Java or Kotlin). Focusing originally on use cases of the financial industry, a Corda ledger supports smart contracts that may contain an associated legal prose (i.e. Ricardian Contracts [14]), providing a root for the legitimacy of the corresponding code. Corda has been designed from the ground up to support a global network of smaller business networks (pattern #3 in Tab. 1) and recently also offers a DID method implementing the W3C standards [15].

---
[1] https://www.corda.net

As a key differentiator to other DLT approaches, the existence of a global state for the entire blockchain is optional (usually not implemented). Corda transactions still are cryptographically linked to the transactions they depend on, but communicated P2P exclusively to involved nodes for processing. Transactions thus are approved "immediately", increasing the throughput of the entire system, but nodes can "see" exclusively those transactions in which they are participating and are consequently limited in their ability to validate a transaction involving user(s) from another subnet. In such cases, additional overhead needs to be spent by inquiries through a specialised "notary" node. Corda networks fit well use cases in environments of highly regulated clusters, like services of the financial industry, where the global network can be broken down into multiple sub-networks.

**MultiChain: Permissions and Predefined Rules**

Forking off the public Bitcoin blockchain, the MultiChain[2] approach has been developed as a private, permissioned network that – as suggested by its name – allows to interoperate multiple parallel, but not hierarchical, instances of the blockchain with cross-chain applications [16]. Technically, the MultiChain approach addresses two general bottlenecks of blockchains: privacy is enabled by "streams" that only are accessible to users with the corresponding privileges, and scalability is improved by an integrated off-chaining strategy (Sec. 1.3). Only the hashes of heavy data are stored on-chain whereas the data itself resides outside the chain, either in form of a (i) centralised repository, or (ii) handled by a P2P file sharing system, or (iii) stored in local repositories that are managed through MultiChain itself.

A central difference of MultiChain lies in the implementation of rules that control transactions: smart contracts are predefined rather than established by the users, which merely may select for each use case a certain level of validation stringency by disabling some of the predefined rules through so-called "stream filters". The stream concept in MultiChain ledgers thus allows users with corresponding permissions to set up additional scopes for groups in addition to the "root stream" of a blockchain (pattern #3 in Tab. 1).

*2.2 Account models for enterprise solutions*

The UTXO paradigm (Sec. 2.1) is well suited for global ledgers with similar transaction types, its limited set of smart contracts hampers a flexible implementation of generic business processes, where transactions are largely independent of each other and do not affect many pieces of information at the global scale. Therefore, enterprise solutions predominantly employ "account" models that bind smart contracts to the data they are controlling within the ledger. By this, the scope of data that can be altered by a piece of chain code, and *vice versa* also the code that is able to modify certain data in the ledger, is strictly defined. Corresponding ledgers are modular systems of atomary databases where data is exchanged in the form of messages to and from smart contracts controlled through the "accounts" of the users. Such messaging often provides a more natural approach than UTXO to model use cases of pattern # 3.

**Quorum: Hybrid Network with Public and Private Transactions**

Developed originally by the investment bank JPMorgan and later-on acquired by the ConsenSys software company, Quorum[3] is an open-source, enterprise-focused, permissioned blockchain that has been developed off the official Ethereum protocol[4]. In contrast to Ethereum, however, Quorum exclusively admits participants to the network who have been pre-approved by a designated authority, the so-called "consortium". This avoids potential threats of catastrophic failure or security breach, and at the same time enables the exchange of private P2P-transactions in addition to the public messaging inherited from Ethereum. Quorum therefore is considered a "hybrid" architecture in the language of [17].

With ZKP enabled technologies available, Quorum is in principle suited to model use cases of any of the four patterns described in Tab. 1. However, the constellation P2P messaging is cryptographically not perfect and under certain circumstances (i.e., with pre-knowledge of the network mapping) nodes can gather some basic activity data about private transactions that they are not involved in [18]. Moreover, Quorum also does not natively provide off-chaining strategies for the private states of the nodes, which can lead to scalability issues when the number of data exchanged privately grows big.

---

[2] https://www.multichain.com

[3] https://consensys.net/quorum/
[4] https://ethereum.org/en/

**Hyperledger Fabric: a Multi-Ledger Network**

Also Hyperledger Fabric[5] (simplified denoted "Fabric") is an open-source, permissioned, private blockchain system, which has originally been initiated by Digital Asset and IBM[6] but now evolved as a collaborative cross-industry venture hosted by The Linux Foundation [19]. To enable privacy, the Fabric architecture divides the network into "channels" and the data into "private data collections". This two-level privacy mechanism provides flexibility in modelling different business policies, but also impacts negatively on the performance, since creating a channel within a channel requires intense computing [20].

Another key difference of Fabric is the "endorsement" approach, which provides a high level of flexibility for the chaincode and still guarantees determinism: before committing a message to some chaincode a peer is required first to send the corresponding message to some endorsing peers for independent execution, to receive their opinion on the safety in form of an "endorsement". However, as all the endorsing and committing peers perform the encryption/decryption operations, the endorsement policy is also costly in computations. In a nutshell, Fabric multi-ledger-based architecture can be flexibly used in use cases involving several collaborating organisations, particularly as outlined by our patterns #2 and #3 (Tab. 1).

*2.3 Layer 2: interoperable blockchains*

Layer 2 refers to a secondary system that sits on top of a blockchain network to relieve some of the workload employing side-chaining, child chaining, or off-chaining approaches. One first step in this direction was the "side-chaining" approach originally proposed by the Elements platform[7] to enhance performance and privacy of the public Bitcoin blockchain [13]. Elements side-chaining employs special nodes called "Watchmen", which move ("peg in") multi-signature transactions for verification from the main blockchain to a separate blockchain, and subsequently back to the main chain ("peg out").

**OpenChain: side-chaining with anchors**

The OpenChain[8] platform developed by CoinPrism later-on leveraged such Watchmen concept for blockchain interoperability employing the Lisk platform[9]. Through Watchmen OpenChain "freezes" cumulative hashes of transactions created in (possibly private "close-loop") side chains to a public main blockchain (i.e. a Bitcoin system) in a process called "anchoring". This allows OpenChain to validate side-chain transactions without adopting the concept of a block, and the responsibility of immutability is delegated through anchors to the public blockchain. However, OpenChain's "side-chains" in fact do not employ blockchains but structure transaction data that can be interlinked employing hash anchors from the main chain.

**Bitcoin Lightning: cryptocurrency side-chains**

The Lightning Network[10] is a system that is built on top of a public blockchain to facilitate fast peer-to-peer transactions. The most popular application of the Lightning Network is in combination with a Bitcoin blockchain, but it constitutes a separate solution and also other cryptocurrencies such as Litecoin have integrated it. Separated from the Bitcoin network, the Lightning Network operates with its own nodes spawning "private channel" networks that can be seen as mini-ledgers for fast P2P transactions that are not dependent on building blocks. Special transactions are necessary on the main blockchain to initiate or end a private channel, involving locking of the assets that can be spent in the P2P side-chain. Moreover, in the Lightning Network the side-chains must be of the same nature as the main blockchain.

**Ethereum Plasma: hierarchical child chains**

Plasma chains[11] in Ethereum are separate child blockchains anchored to the main Ethereum blockchain by Merkle trees. In principle, each Plasma child chain can be interpreted as a customisable smart contract that is designed to serve particular needs. The communication between the child chains and the root chain is secured by so-called "fraud proofs", which delegate to the root chain the responsibility of keeping the network secure and of propagating malicious actors across child chains. The tree structure allows the anchoring of child chains also cascaded in hierarchies, but they all are required to run the same blockchain technology in order to guarantee compatibility with the main Ethereum blockchain.

---

[5] https://www.hyperledger.org/use/fabric
[6] https://www.ibm.com/topics/hyperledger
[7] https://elementsproject.org

[8] https://www.openchainproject.org
[9] https://lisk.com
[10] https://lightning.network
[11] https://ethereum.org/en/developers/docs/scaling/plasma

### 3. The OpenDSU Platform

Although recent layer 2 approaches allow operating multiple blockchains of the same or very similar DLT to distribute the workload, mainly of financial transactions in cryptocurrency networks, our elaborations in Section 2 demonstrate that currently there is no platform to orchestrate multiple, arbitrarily different DLTs in a single enterprise environment. Going back to the PrivateSky Project[12], we therefore sought to develop a platform we call "OpenDSU" [21], which is to fill in this gap. Since then, our OpenDSU SDK[13] matured and improved by employing it in research and commercial projects - such as the currently ongoing PharmaLedger[14] Initiative (Sec. 4). The vision of OpenDSU is to provide a framework for combining blockchains and other DLTs, existing ones such as the ones developed in the future, without any pre-requirement on their architecture and without compromising performance and privacy (Fig. 1).

Some of the key contributions by OpenDSU are:

- *Digital sovereignty*: the OpenDSU concept was designed from the ground up to allow nodes with SSIs from different so-called blockchain domains, intrinsically linked to a finely granular permissioning system (Sec. 3.5).
- *Enterprise smart contracts*: in the spirit of the smart contract concept outlined in Sec. 2.2, OpenDSU binds chain code to the data it controls, in containers called DSUs ("data sharing units", Sec. 3.2).
- *off-chaining*: based on these DSUs, OpenDSU introduces a universal off-chaining strategy (Sec. 1.3) that allows storing of so-called *near*-chain data across different blockchain domains as "bricks" in "bricks storages" (Sec. 3.3).
- *Multi-chaining:* starting off from the idea "*no size fits all*", OpenDSU leverages anchoring techniques (Sec. 2.3) to bind near-chain data to possibly hierarchic ledger network(s), shaping a technology-agnostic, multi-chain platform (Sec. 3.3).
- *Off-the-shelf platform*: the source code of all OpenDSU components is transparently available and can be employed "out of the box", without any vendor lock-in restrictions.

---

[12] https://profs.info.uaic.ro/~ads/PrivateSkyEn/
[13] https://opendsu.com/
[14] https://pharmaledger.eu/

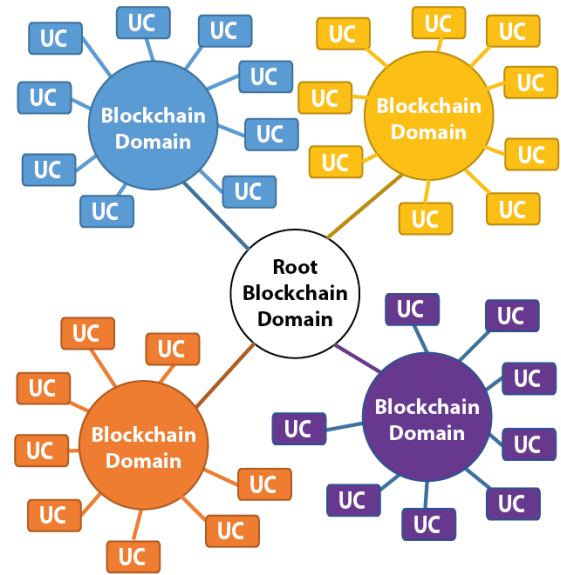

*Fig. 1: Blockchain Domains.*
*The Open-DSU approach orchestrates a multi ledger network across different blockchain domains, each one hosting possibly multiple use cases (UCs).*

In the remainder of this section, we present in more detail the key components of our OpenDSU platform [22].

*3.1 Overview and architecture of the OpenDSU platform*

Fig. 2 sketches the logical layout of any OpenDSU platform. From a technical point of view, all software components can be segregated into three structured layers, i.e., the network (layer 0), the on-chain (layer 1), and the "near"-chain layer (layer 2):

**Layer 1: the ledger network(s)**

The centre panel of Fig. 2 shows the actual DLT networks, which we consider the OpenDSU "Layer 1", each one running several replicas. In OpenDSU we have support for Quorum (Ethereum) blockchains [23], Hyperledger Fabric blockchains (experimental) [24,25], or instances of our OpenDSU BricksLedger (Sec. 3.3) network [21,26–28]. Further DLTs may be added in the future, requiring merely a correspondingly implemented and properly configured *adapter*.

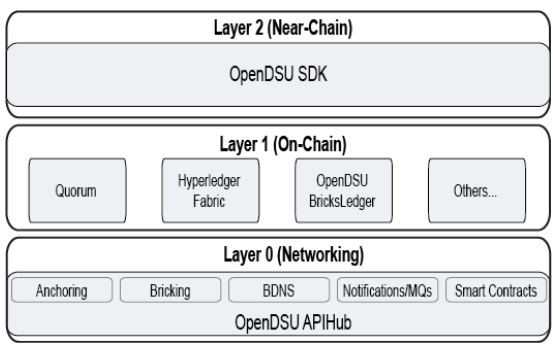

*Fig. 2: Layers of an OpenDSU platform.*
*Bottom: Layer 0 is the networking layer that provides the interconnectivity between the replicas of the blockchain domain(s). Centre: Layer 1, the "on-chain" layer, provides possibly various ledger technologies with their respective consensus-building methods. Top: Layer 2 consists mainly of the OpenDSU SDK library and manages the near-chain processing.*

In our OpenDSU paradigm, the primary purpose of blockchain(s) is to work as notarization mechanisms for DSUs, allowing the integration of different blockchains that may provide, for instance, different security models and/or scaling capabilities. To this end, OpenDSU has been conceived from the start to be *agnostic* towards the employed blockchains. In order to enable the interoperability of heterogeneous and not *a priori* fixed DLTs, OpenDSU provides an independent naming system we call BDNS (*Blockchain Domain Naming System*) [29].

BDNS is essentially a discovery mechanism that translates numerical addresses needed for locating and identifying computer services and devices of the underlying network protocols to more easily memorizable domain names in the form of "*subdomain.rootdomain.topdomain*". In this light, BDNS is a service for blockchain bootstrapping by trusted configurations, similar to DNS, DPKI, and other forms of verifiable mapping, which however in contrast to the latter aims to be a secure, hierarchical, decentralised and self-sovereign naming system [29]. BDNS enables smart contracts on hierarchical blockchain domains, agnostic to the respective DLT employed [30].

**Layer 2: the OpenDSU SDK**

Layer 2 (Fig. 2, top panel) consists of the OpenDSU SDK, a collection of high-level functions associated with the "blockchain domains" endpoints, specified by an hierarchical path in the form of *"component > blockchainDomain > action > parameters"*. Whereas the traditional nomenclature for blockchain systems usually distinguishes merely data stored in the ledger (*on-chain*) from *off-chain* data that can be controlled independently, we further discriminate three classes of the latter:

(*i*) *near-chain* data is wrapped into so-called DSU containers (Sec. 3.2), which can be exported and validatably reimported to and from external premises employing "bricking" and anchoring techniques (Sec. 3.3).

(*ii*) *far-chain* data in OpenDSU is exported from the ledger without anchoring to a blockchain, e.g. by employing databases [31].

(*iii*) *no-chain* data is used by the OpenDSU platform, but not employed by the DLTs in Layer 1 at any time.

This refined off-chain policy allows us to differentially outsource from the blockchain such shared data, which needs consistently to be accessed by different stakeholders (near-chain) as well as sensitive data that often follows business-specific policies (far-chain).

**Layer 0: the APIHub**

The networking layer or Layer 0 (Fig. 2, bottom panel) comprises different components (e.g., BDNS, bricking, anchoring, etc.) running in the so-called *APIHub*, the backend of our OpenDSU platform. The APIHub acts as a server that offers – typically to wallets – different functionalities provided by the OpenDSU SDK (Layer 2). According to their specificity for a DLT, we further distinguish *adapter-components* relying on specifically implemented blockchain adapters (e.g., BDNS) from adapter-independent components (e.g., bricking). The OpenDSU APIHub is extensible, i.e. new components may be added in custom implementations. In our present OpenDSU implementation, all APIHub computations are being executed in a single/main thread, but our ongoing efforts comprise the development of a multi-threaded APIHub to increase performance of the OpenDSU platform.

*3.2 Sharing data with DSUs*

**Simple DSUs**

In OpenDSU – as suggested by the name – data sharing units (DSUs) are central components, which we designed from the start as *near-chain* containers [32]. When exported to external storages (Sec. 3.3), DSU contents are encrypted and tokenized to provide confidentiality and privacy. They can only be decrypted and re-assembled to a DSU after a suitable access key (a "KeySSI", Sec. 3.4) has been resolved by a server-, cloud- or edge-based wallet. Once the so-called *KeyResolver* resolves a KeySSI to a certain DSU, the corresponding near-chain data – provided the necessary access level – is decrypted and loaded in the execution environment.

Fig. 3 (upper panel) shows that, after assembly, a DSU instance becomes available in the corresponding execution environment, usually a sandboxed container [33]. A DSU spawns a micro file system that also can be interpreted as a key-value micro-database, with the path of each file representing the key pointing to the file's content. DSUs can contain both, *near-chain* data and also chain-code (i.e., smart contracts), which is not visible on-chain and therefore can also be considered a form of "*secret"* smart contracts.

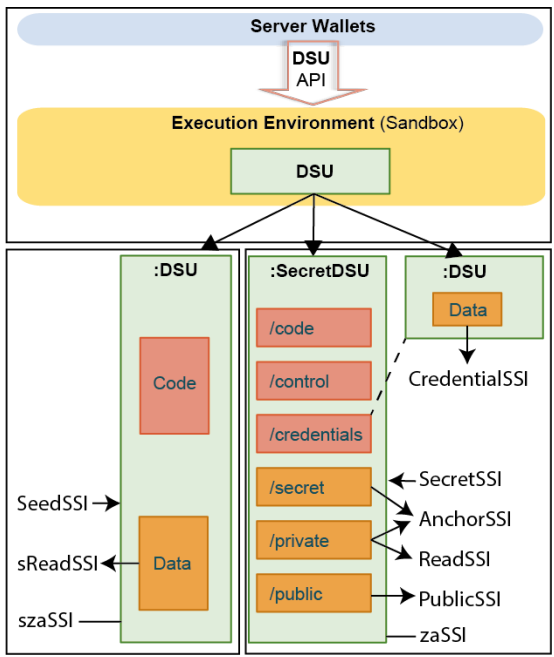

*Fig. 3: Overview of the DSU core concept.*
Lower-Right: Lower-Left: SecretDSUs employ the SecretSSI family of KeySSIs with the derivation hierarchy: SecretSSI (DSU ownership) > AnchorSSI (writing) > ReadSSI (reading) > PublicSSI (basic DSU access) > zaSSI (reference without access).

Fig. 3 (lower left panel) outlines the finely granular access control to DSUs based on so-called *families* of keySSIs, created from one another by "derivation" (Sec. 3.4). For instance, a KeySSI family popularly employed with DSUs that are shared only amongst a rather limited set of users starts off with the generation of a private key that provides full control (i.e., ownership) of the DSU, the *SeedSSI*. From this, a hashed version called *"sReadSSI"* (*seed read SSI*) then is "derived" to be shared amongst the group of users who encrypt and decrypt the collective data. Further derivation produces a *"szaSSI"* ("*seed zero access"*), a public key that provides no access to the DSU itself but exclusively to a list of hashes representing the history of previous versions of the DSU, suitable for signing and/or anchoring data (e.g., digital wallets).

**Combined and complex DSUs**

Obviously, a DSU may contain as data one or more KeySSIs that resolve to other DSU instances. In this light, we propose the design pattern of *constSSIs*, i.e. KeySSIs in a human-readable format that resolve to an immutable, read-only, "dummy" DSU containing a cryptographically more secure SSI that contains random numbers and therefore cannot be memorised intuitively anymore. The constSSI pattern circumvents a problem known as "Zooko's Triangle", according to which identifiers of any naming system cannot achieve all three attributes at the same time: meaningfulness to human, cryptographic security, and decentralisation (i.e., self-sovereignty) [34]. Apart from pointing to each other by resolving contained KeySSIs, the DSU API also allows users to dynamically "mount" DSUs into one another, aggregating their contents but preserving their inherent access control. This enables a generic creation of nested containers that provide custom access levels across a possibly heterogeneous group of users.

For common use cases, OpenDSU additionally provides an already predefined class of "SecretDSUs" with a list of standard folders that can be accessed through different KeySSIs from the *SecretSSI* family: a "code" folder with the mounted DSU type; a "control" folder containing a whitelist of public keys that can modify the DSU, serving as an additional protection against attacks from trusted groups in possession of an *AnchorID*; a "public" folder containing the public key that provides basic access for holders of *PublicSSI* (or above) privileges; a "private" folder with confidential data readable only when employing a *ReadSSI*; a "secret" folder with private keys, e.g. for anchoring (i.e., *AnchorSSIs*); and an externally mounted "credentials" folder accessed using a *CredentialSSI*, with the signatures employed to validate versions of this DSU. As can be seen, *CredentialSSIs* do not stem from the SecretSSI family (Fig .3, lower left panel).

*3.3 Near-chain data management*

**Brick Storages**

As introduced in the previous section, the contents of a DSU are assembled and decrypted dynamically from their persistent storage form, so-called "brick storages" [35]. Physically, a brick storage can store data on virtually any storage medium, e.g., locally, remotely, or in the cloud. Technically, a brick storage is a simple web service capable of storing and retrieving bricks for clients (i.e. wallets) that know the identifying brick hash. Due to obvious security motivations, DSU data needs to be encrypted when exported to a brick storage. OpenDSU employs symmetric encryption, each brick is decrypted with its own particular KeySSI (Sec. 3.4).

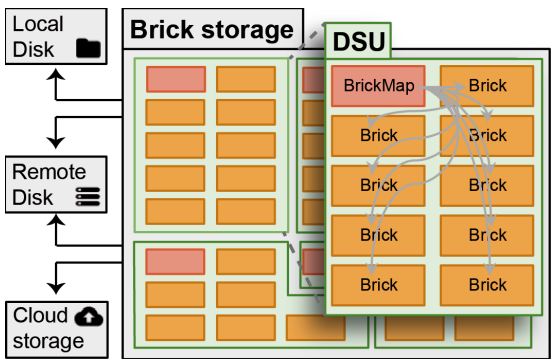

*Fig. 4: Brick Storage.*
*Schema of a brick storage with bricks (orange boxes) and brick maps (red boxes) storing the DSU information off-chain either locally, remotely, or in a cloud.*

Fig. 4 outlines the brick representation of DSUs. To prevent undesired monitoring of changes, DSUs are disassembled into different bricks, in a process we call "bricking". Special bricks, that we refer to as "brick maps" – respectively "bar maps" in the legacy terminology of the earlier PrivateSky reference implementation of OpenDSU [36] – store the KeySSIs for the bucket of bricks they are referencing. As sketched in Sec. 3.1, a DSU during its life cycle is subject to changes, which may either alternate the content and thus also the hash identity of the DSU, or remove and respectively delete the DSU entirely. Despite these possibly high dynamics in modifications on a DSU, a brick map once created remains unchanged and referential by its hash in the persistent off-chain brick storage. Therefore, a new version of a modified DSU also requires a new brick map, and we refer to the list of all brick maps since the creation of a DSU as the corresponding DSU *history*.

Based on the bricking mechanism, we also developed *BricksLedger*, an OpenDSU component that is able to transform any database or blockchain into a ledger appropriate for OpenDSU anchoring [37]. On the one hand, BricksLedger provides a convenient drop-in for the use of standard databases or other existing data warehouses together with other ledgers in OpenDSU, achieving blockchain level audit for corresponding data. On the other hand, the possibility to anchor DSUs to virtually any data storage can effectively mitigate concerns about confidentiality of on-chain transactions and about performance (i.e., speed/throughput of the ledger).

**Anchors**

Calls to the DSU API (Sec. 3.2), e.g. writing files, keys and complementary functions, may modify the content (i.e., the "state") of a DSU, creating a new version of the DSU with a new ID that represents its public key. DSU modifications are made persistent by writing a new anchor to the blockchain, which comprises a hash code combining the new public key of the DSU as well as the DSU's history. Importantly, any write operation to a DSU will be observable by other execution environments only *after* the new version is successfully anchored, and all previous versions of a DSU are administered by a fully resolved history, signed by corresponding hashes.

The schema in Fig. 5 outlines that OpenDSU employs an anchoring technique to irrevocably bind DSUs to a blockchain, enabling the integrity and traceability of data, e.g., to verify near-chain data from a brick storage. An *anchor* is composed by an identifier (i.e., *AnchorID*) – i.e., a special type of KeySSI identifying the DSU – and the *history* of brick maps from the respective DSU, represented by hash links also in the form of special KeySSIs. By this, OpenDSU anchors are self-validating, i.e. the DSU history of hash links in an anchor can only be truncated but not altered by manipulation. Anchoring therefore allows for digitally signing data and code – in its initial version and all subsequent versions/updates [28].

After thorough research of the anchoring problem, we subdivide anchors into at least two types: implicit and explicit anchors. *Implicit anchors* are created intrinsically as a result of blockchain operations that are not explicitly stored in the world state (i.e., cache) of OpenDSU. To this end the immutable nature of blockchains serves as a notarization mechanism for events or other specific information requiring "notarization". However, OpenDSU avoids the usage of implicit anchoring by additionally providing *explicit anchors* implemented as smart contracts or in the form of some services outside of the ledger. By explicitly storing history and timestamps, explicit anchors are easy to reconstruct and to validate, as defined by anchoring smart contracts.

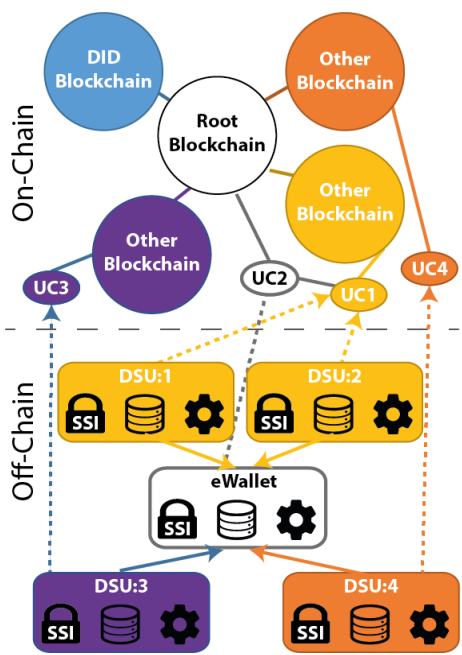

*Fig. 5: Anchoring in OpenDSU.*
*Blockchain anchors (lines) between different ledger domains (top panel) are resolved by the BDNS system (Sec. 3.1). Near-chain data is organised into DSUs (bottom panel), that can be mounted (solid arrows) into each other (Sec. 3.2). Anchors (dashed arrows) binding DSUs on the left to ledgers to the right enable the verification of near-chain data.*

Explicit anchors are further divided into heavy and light anchors. Heavy anchors ensure that anchoring is controlled and handled exclusively in smart contracts on-chain. In ledgers with support for smart contracts, they provide a costly though flexible solution considering the possibility of custom validations, the persistent hash history, and the timestamps. Light anchors, in contrast, require the on-chain storing of merely a list of hashes and a pair of anchor identifiers, obtained through zero access KeySSIs (Sec. 3.2). Anchored data as well as its history are in this case both reconstructed and validated from off-chain storages [38]. As an example for the philosophy behind our different anchoring strategies, we sketch here two ways of implementing Zero Knowledge Proofs (ZKPs) in OpenDSU: on the one hand, ZKP protocols are traditionally implemented with implicit anchors, relying on the cryptographic properties of a blockchain. On the other hand, hash links could be implemented as explicit light anchors, accompanied by some custom cryptography to obtain ZKP values that can be used to prove computational integrity properties (i.e., zero access anchoring). Following the philosophy of not basing the privacy properties of a use case on a new and often not fully proven cryptography, the anchoring control in OpenDSU has been specifically conceived for light anchors, typically employing digital signatures. In use cases with an anticipated high number of anchors, light anchors also allow designing specialised high-throughput blockchains storing, e.g., billions or even trillions of light anchors (Sec. 4).

*3.4 KeySSIs*

**Self-sovereign IDs across ledgers**

The DID (Sec. 1.2) community currently explores possibilities to overcome shortcomings in existing wallet technologies by developing novel methods for the communication between wallets that are capable of receiving credentials but also of transmitting and presenting VCs. A step ahead in this direction constitutes the DIDComm[15] approach, which relies on DIDs that do not require blockchain anchoring, ensuring a very high level of privacy [39]. However, W3C protocols like the DIDComm approach cannot be integrated well with our OpenDSU concepts. Firstly, they have been developed for home users and impose a rather unnecessary and complex overhead when implementing enterprise solutions (Sec. 4.1). Secondly, these anchor-less DID approaches cannot employ our BDNS functionality (Sec. 3.1). Although not mandatory in OpenDSU, KeySSIs are typically blockchain- anchored, enabling self-validation [40,41]

For OpenDSU, we therefore designed a blockchain agnostic through anchored SSI identifier concept through *KeySSIs*. As indicated by their name, KeySSIs are bifunctional: they can be used as *keys* for de-/encrypting (parts of) the information in DSUs or in bricks (Sec. 3.2), and at the same time they represent *identifiers* of their owners, who can be of physical (i.e., individuals), juridic (e.g., companies) or technical nature (e.g., resources/processes). Access privileges held by a certain KeySSI further can be delegated – entirely or in parts – to other KeySSIs

---
[15] https://didcomm.org/

*derived* from it, which subsequently are subtypes accompanying the original KeySSI. There are many possibilities to create KeySSI types/subtypes and we classify these into so-called "families".

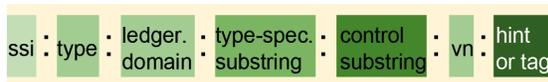

```
ssi:seed:ePI.pharma:RANDOMSEEDKEY:HASHRANDOMKEY
ssi:za:ePI.pharma:HASHSERIALISATION:HASHPUBLICKEY
```

*Fig. 6: Syntax of KeySSI Identifiers.*

Fig. 6 summarises the structure of KeySSI descriptors, which is inspired by the W3C DID standard but provides an extended syntax to describe additional parameters as required in the OpenDSU system. In a nutshell, a very unique aspect of the KeySSI specification is the ledger/blockchain domain ($3^{rd}$ field) to which the SSI belongs, enabling an identity system that spawns multiple, possibly hierarchically cascaded ledgers respectively blockchains through BDNS (Sec. 3.1). The "type" field, as an equivalent to the DID "method" string, describes the specific (sub-)types of KeySSI families in OpenDSU. A complete explanation of all KeySSI field descriptors is provided by Appendix A.2. Like the DIDs proposed by the W3C, KeySSIs can be controlled in a decentralised manner by possibly multiple domains. In this regard DIDs and KeySSIs can be considered two different methods for enabling SSIs (Sec. 1.2).

**KeySSI-based messaging through portable smart wallets**

Sec. 1.1 outlines the progress of developments around web3, requiring ever more types of information in a digital wallet that cannot be provided by centralised wallet services. For instance, the currently popular Apple Pay[16] and Google Pay[17] wallets allow rather limited information to be "carried" inside them, e.g. credit cards, tickets for travelling or events, and other basic credentials, and cannot cope with DIDs or SSIs. The ability to flexibly represent user identities and at the same time to act as cryptographic keys opens up several possibilities of employing KeySSIs, and in order to overcome aforementioned shortcomings in existing digital wallets.

In OpenDSU, we therefore employ KeySSIs to implement a more simple yet effective protocol for communication between wallets. This OpenDSU messaging protocol employs the ECIES (Elliptic Curve Integrated Encryption Scheme) [42,43] to encrypt messages sent between DIDs based on KeySSIs in different scenarios of messaging [44]. By this, KeySSIs can be exchanged to provide access to "messages" in the form of DSUs to which they resolve. To allow the exchange over unencrypted channels, we propose specifically encrypted *encSSIs* as wrappers for the exchanged KeySSIs [45].

In addition, we employ AnchorIDs in notifications as names of (public) channels to provide "addresses" for DSUs that act as a mailbox (i.e., "message queues, MQs"): everyone can write to a MQ, but only the respective owner can read the corresponding messages [46]. MQs can also be employed for automatic updates, when new versions of a DSU are anchored. Ongoing efforts aim at providing a generalised programming model, based on the concept of executable choreographies [47–49]. In this regard, the keySSI based communication between digital wallets in OpenDSU can provide a platform for developing modern protocols obeying the principles of privacy and security that may replace the e-mail system as it existed since the dawn of the internet. Such protocols could be designed from the beginning to be decentralised and difficult to centralise, overcoming current problems of monopolisation in the area of email services (Sec. 1.1).

## 4. Implementation and Optimisations of OpenDSU in PharmaLedger

Over the recent years, blockchain based initiatives have emerged all over the world to leverage digital sovereignty across different sectors, such as government, economics, energy and health [50]. The OpenDSU SDK library [22], originally implemented by the PrivateSky project[18], has been improved by insights from different research as well as commercial projects, e.g. in the currently ongoing PharmaLedger Project[19]. In the remainder of this section, we explain the PharmaLedger contribution to the OpenDSU open-source project [51], including use case validations, code improvements, and proposals for further improvement.

PharmaLedger is a multinational initiative, funded by the European Community and by IMI industry partners, to address several challenges of exchanging information in the pharmaceutical and in the health sector: sharing of patient records, communication between partners in the drug and medical equipment supply, etc. The PharmaLedger Project therefore includes public authorities, medical partners from the public sector, and commercial enterprises such as software companies and marketing authorization holders (MAHs) of the pharmaceutical industry. The heterogeneity in business policies of the different stakeholders renders PharmaLedger an excellent opportunity to evaluate the potential of our OpenDSU platform.

---

[16] https://www.apple.com/apple-pay/
[17] https://pay.google.com/
[18] https://profs.info.uaic.ro/~ads/PrivateSkyEn/
[19] https://pharmaledger.eu/

The initial working package of PharmaLedger prioritised, from an initial list of >100 use cases, seven use cases to be implemented within the pilot phase of the project, categorised into one of three so-called "domain reference applications" (DRAs):

• DRA1 – "supply chains": e.g., the *pharmaceutical supply chain* (supply chain) use case, for tracking all the pharmaceutical items produced by each of the MAH partners, with an estimated number of up to billions of transactions per day;

• DRA2 – "health data": e.g., the *electronic product information* (ePI) use case providing the MAHs' leaflets with information on their pharmaceutical products in an updatable form that granularly can be approved by the health authorities;

• DRA3 – "clinical trials": e.g., the *electronic patient consent* (eConsent) use case that provides patients to dynamically provide respectively revoke by smart contracts their agreements on the use of their data in clinical studies;

*4.1 DIDs in PharmaLedger*

In an early working package of the PharmaLedger Project, we investigated the possibility of employing the DIDComm and other existing approaches to SSIs, based on the standards proposed by the W3C DIDs (Sec. 3.4). In a nutshell, our experiences led us to the conclusion that these standards have been developed for home users, who mostly aim at total anonymity and non-correlatability. However, the challenges of providing security and confidentiality of systems in an enterprise environment differ substantially from these issues of privacy protection.

More precisely, technologies along the lines of DIDComm optimise for enterprise environments that rely on VCs for creating trust. Consequently, issuers taking the role of a "root of trust" conceptually do not differ much from the approach of X.509 certificates. Although their more modern syntax and additional extensions may be beneficial from the confidentiality point of view, they also create an unnecessarily high degree of complexity, especially when considering ZKPs. Existing enterprise solutions for DIDs thus focus on circumstances that actually do not exist in a Digital Trust ecosystem such as PharmaLedger. We therefore estimated that adopting the current DIDComm approach to the real-world demands by the different PharmaLedger participants would infer an overhead that exceeds the concrete benefits.

Therefore we decided to employ keySSIs and the BDNS concept of OpenDSU. This pragmatic approach allows us to adapt and integrate heterogeneous technical solutions for identifying patients, or citizens, across countries. Of note, OpenDSU wallets will still be able to adopt DIDComm or similar technologies along the same lines in the future, when standards have matured to overcome current limitations. The challenge of inter-country barriers caused by differences in the legal preconditions for health data policies is a paramount precondition for several PharmaLedger use cases, and has recently been leveraged by the General Directorate for Health and Food Safety of the European Council publishing a proposal on the regulation of the European Health Data Space [52].

*4.2 APIHub improvements for Pharmaledger*

The PharmaLedger OpenDSU architecture still includes the OpenDSU APIHub on the Network Layer 0, a blockchain technology that achieves consensus (Layer 1) and OpenDSU SDK in DSU Layer 2 (Sec. 3.1). However, as the APIHub constitutes the main backend component for all use cases of the OpenDSU platform, it is instantiated in PharmaLedger separately for each use case. Furthermore, when benchmarking relatively early in the project an implementation of the ePI use case employing the Quorum blockchain[20] with our OpenDSU APIHub, we identified three major shortcomings of the original PrivateSky BricksLedger component (Sec. 3.3) that needed further improvement:

*#1 Latency*: a call on the Quorum blockchain takes a few seconds to be answered[21], s.t. a sequence of 2-3 calls, as required by most business use cases, causes delays of 5–10 seconds until the confirmation of a transaction and leads to a poor user experience.

*#2 Throughput*: the scalability of the Quorum blockchain is limited to a few hundreds of transactions per second [29].

*#3 Security*: custom PPP-based protocols for the communication of blockchain replica (i.e. DevP2P[22])

---
[20] https://www.kaleido.io/blockchain-platform/quorum
[21] https://consensys.net/quorum/
[22] https://github.com/ethereum/devp2p

move MAHs to rely on weak security models, e.g. by employing a single cloud provider to host all blockchain nodes in a virtual network.

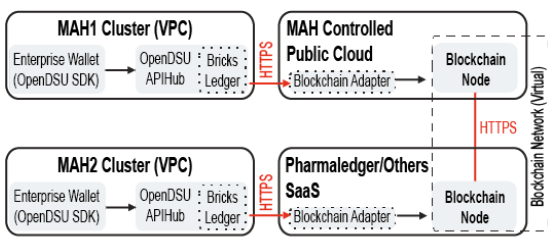

*Fig. 7: Deployment Clusters.*
*Red arrows indicate the HTTPS communication paths between APIHub and blockchain nodes.*

To achieve the trustless aspect of blockchain technologies, challenge #3 motivated us to have two deployment clusters: one for corporate internal transactions of MAHs, and another one spawning a blockchain network across consortial or public services. By this, each MAH needs to deploy and control a separate APIHub, requiring deployments different from the blockchain node. Like in the communication between blockchain nodes in OpenDSU, both "clusters" of APIHubs communicate via standardised and encrypted HTTPS, eliminating concerns about non-compliance with company security policies (Fig. 7).

*4.3 Optimistic execution of Smart Contracts*

We spent further efforts on researching solutions to mitigate challenge #1 (latency), while improving respectively not compromising challenges #2 and #3 (scalability and security, Sec. 4.2). This led us to conceive an heuristic approach we call the "optimistic execution" of smart contracts [53,54], which enables us to boost the performance of anchoring DSUs in PharmaLedger. The idea for this novel concept arose from our observations of situations in PharmaLedger, predominantly matching use cases of pattern #2 or #1 (Sec. 1.4). Corresponding smart contracts may be executed in an "optimistic" way, i.e., in a way that is not immediately validated by the consensus algorithm.

Fig. 8 shows a layout of our concept of "optimistic execution", which attempts to alleviate the overall network overhead by reducing the workload for reaching a global consensus and instead resolving consensus straightforwardly in the "local" context of anchors. Our heuristic on executing such smart contracts "optimistically" relies on the fact that, in the absence of notarization, it is possible to straightforwardly implement "self-consistent" anchoring commands: the OpenDSU anchoring command implicitly implements a "nonce-based mechanism", and each anchoring operation contains a signature binding it to the request with the last anchor variant, which *per se* is trusted to be correct by the signer.

Groups of anchors can more easily be trusted, especially when owned by a single owner of the corresponding DSU (use case pattern #1 and #2): to our observations in PharmaLedger use cases, most anchors are controlled by a single "producer" actor providing and updating (i.e., writing) data shared with "consumers". Therefore, "optimistically executed" smart contracts also are largely safe against common threats such as "replay" and/or "double spending" attacks. Moreover, since optimistic execution is by design intended to occur mostly in the nodes controlled by the owner of the respective anchors, also a possible issue by partitioning of the network is not a real issue in practice.In any case, implementing a *validated execution*, i.e. an execution that updates the validated world state as in the usual blockchain, will directly eliminate any potential issues.

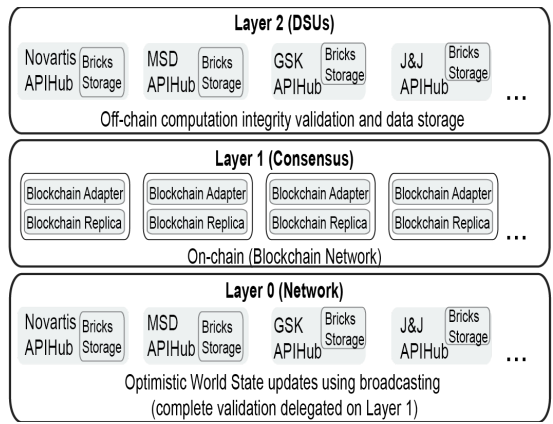

*Fig. 8: Logical view of the OpenDSU layers in PharmaLedger.*
*Validation is delegated from the MAHs' APIHubs to the blockchain nodes (Layer 1).*

In summary, the main advantage of the proposed optimistic execution is that we can answer the clients' requests very fast because the chances of consensus invalidating the commands are zero in most cases. Additionally, the overall performance and scalability of the blockchain will be improved by using this approach for anchoring.

**Discussion and Conclusions**

In the present paper, we introduce the concepts and components of OpenDSU, a generic platform that we developed over the past years and that enables the seamless integration of SSIs, DLTs and off-chain storage in enterprise environment use cases. The benefits of our platform are rooted in the design of conceptually sound components around central data sharing units (DSUs) that shuttle data and code to the endpoints where it is needed. By our universal BDNS interface, OpenDSU is agnostic to the respective technologies employed for spawning ledger networks, i.e. our platform supports the integration of arbitrary DLTs, including but not limited to blockchains.

Notwithstanding this flexibility of integrating even heterogeneous ledger technologies, OpenDSU provides a unified and seamlessly integrated concept for near-chain data, i.e., blockchain data exported to external premises that can be fully validated upon re-import to the ledger. We achieve this through bricking and anchoring, where bricks are atomary data blocks that can be stored in a privacy-preserving fashion on virtually any external medium, organised by blockchain-integrated anchors. Anchors coordinate the decomposition/reassembly of DSUs to/from bricks, maintaining previous versions of the DSU, which can be considered micro-ledgers with full access to their history of modifications.

Outside the execution environment where they are needed, all contents of a DSU are cryptographically secured through symmetric encryption and can be accessed only by employing an appropriate KeySSI. KeySSIs couple sovereign identity management with access control and can hierarchically be derived from one another, delegating some of the privileges from a higher to a lower access level. This combination further enables (automated) messaging through KeySSI identifiers, providing the completely self-sovereign and cryptographically secure exchange of information. Since KeySSIs are agnostic to the DLT(s) they are employed to, they pave the way for powerful cross-DLT applications in the future.

Recently, we successfully implemented our OpenDSU platform in the PharmaLedger Project, which brings together companies and public stakeholders from sectors as different as IT, healthcare and the pharmaceutical industry. As OpenDSU is conceived to employ generic components, without any lock-in model for the pre-defined components, we first conducted thorough research on the inclusion of existing solutions. Regarding the decentralised identity management, however, we pinpoint severe drawbacks by needless overhead and security concerns when employing current DID approaches to the heterogeneous business policies in PharmaLedger, because they have primarily been conceived for home users. Employing KeySSIs through the BDNS of OpenDSU in PharmaLedger allows the development of substantially more flexible and scalable solutions to enable digital sovereignty across multiple ledger domains.

Conversely, our investigations on integrating existing blockchain technologies like Quorum in PharmaLedger also provided insights for improving OpenDSU components. To prevent security concerns that may arise from enterprise-internal data management, we distributed our backend across multiple APIHub instances, with specifically deployed APIHubs dedicated to the internal data management of each company. Our multi-APIHub approach also permits setting up a separate APIHub for each use case in order to segregate the workload by high-throughput scenarios from the remaining operations. To additionally improve the latency and throughput of the system, we also developed an heuristic approach we call "optimistic execution" of smart contracts that resolves the consensus between a limited scope of participants more efficiently in a local context.

The conclusion of all our ongoing research is that we are still only at the beginning of conceiving such future generation ledger systems. In this direction, we developed OpenDSU as a flexible platform to integrate largely arbitrary components for running ledger networks, blockchain anchoring, and the management of decentralised or self-sovereign identities. In the absence of lock-in mechanisms, OpenDSU is promoting a "Darwinian evolution" of the protocols rather than enforcing the standardisation of premature solutions. We conceived all OpenDSU components with the goal of enabling a broader acceptance of ledgers in the long run, by leveraging the full potential of technologies for digital sovereignty.

**Availability**

OpenDSU is freely available under the MIT open-source licence[23] at https://opendsu.com

**Acknowledgements**

Fig. 5 employs three pictograms that may be redistributed under the creative commons, i.e. a lock symbol (CC BY-SA 3.0), a database symbol (CC BY-SA 3.0), and a gear symbol (CC BY 4.0).

---

[23] https://opensource.org/licenses/MIT

According to the ShareAlike agreement, the figure therefore is to be redistributed under the corresponding licences.

# Appendix A

*A.1 List of Abbreviations and Acronyms*

| Abbr. | Full-Form | Section |
|---|---|---|
| BDNS | Blockchain Domain Naming Service | 3.1 |
| DAO | Decentralised Anonymous Organisations | 1.1, 1.3 |
| DLT | Distributed Ledger Technology | 1.1, 1.3, 2 |
| DID | Decentralised IDentity | 1.2 |
| DoS | Denial of Service | 1.3 |
| DSU | Data Sharing Unit | 3.1 |
| DeFi | Decentralised Finance | 1.1, 1.3 |
| DNS | Domain Name System | 3.1 |
| DPKI | Decentralized Public Key Infrastructure | 3.1 |
| ECIES | Elliptic Curve Integrated Encryption System | 3.6 |
| MAH | Marketing authorization holder | 4 |
| NFT | Non-fungible token | 1.1, 1.3 |
| P2P | Peer-to-peer | 1.4 |
| SC | Smart Contract | 4.3 |
| SSI | Self-Sovereign Identity | 1.2 |
| VC | Verifiable Credential | 1.2 |
| ZKP | Zero-Knowledge Proof | 1.1, 1.3 |

*A.2 SSI Structure*

Field descriptors of the KeySSI syntax:

(1) Schema Identifier: "ssi" for keySSIs (cf. "did" in the W3C standard for DIDs)

(2) Equivalent to the DID "method" string, KeySSIs are also categorised by a string describing their (sub-) "type.". SSI (sub-)types are compatible with each other and with the W3C DIDs, which allows a standard KeySSI resolver to implement any of these types.

(3) In addition to W3C DIDs, the SSI syntax provides a unique field to identify a blockchain or "ledger domain". Along the lines of the DNS system in the internet, OpenDSU introduces the "Blockchain Domain Naming System" (BDNS) to provide intelligible names for (sub-)networks, computers, end point services and users in – possibly hierarchical (Section 3.3) – blockchain networks [29].

(4) The "type-specific substring" simply provides sufficient random bits for good security attributes.

(5) The "control substring" specifies the type- (2) specific algorithm used by anchoring services for validating and verifying requests for a new version of the anchored DSU.

(6) The "vn" string is optional (defaults to "v0") and informs the version number *n* of the KeySSI type (2). Not to be confused with the DSU versioning, a KeySSI type version specifies the cryptographic primitives and conventions, e.g. the hash functions and other methods to be used by the anchoring service [37]. New KeySSI type versions must be approved by the OpenDSU standardisation body, providing corresponding RFC documentation.

(7) The (optional) "hint or tag" string provides a generic way of extending KeySSIs for several purposes, again referring to the given (sub-) type (2). It typically "hints" at additional information for the KeySSI resolver, e.g. a favourite server proposed by the owner of the KeySSI. Alternatively, this string can be employed as a "tag" to mark KeySSIs for specific purposes; for instance DSUs that contain sensitive information will require additional data protection mechanisms in place.